# Statistical Detection of Adversarial examples in Blockchain-based Federated Forest In-vehicle Network Intrusion Detection Systems

Ibrahim Aliyu[a], Sélinde van Engelenburg[b], Muhammed Bashir Mu'azu[c], Jinsul Kim[d*], Chang Gyoon Lim[a*]

[a]Department of Computer Engineering, Chonnam National University, Yeosu, 50 Daehakro, Jeonnam 59626, South Korea
[b]Faculty of Technology, Policy and Management, Delft University of Technology, the Netherlands
[c]Department of Computer Engineering, Ahmadu Bello University, Zaria, Nigeria
[d]Department of ICT Convergence System Engineering, Chonnam National University, Gwangju,, Korea

*(Corresponding Author: Email: cglim@jnu.ac.kr, jsworld@jnu.ac.kr)

**Abstract**

The internet-of-Vehicle (IoV) can facilitate seamless connectivity between connected vehicles (CV), autonomous vehicles (AV), and other IoV entities. Intrusion Detection Systems (IDSs) for IoV networks can rely on machine learning (ML) to protect the in-vehicle network from cyber-attacks. Blockchain-based Federated Forests (BFFs) could be used to train ML models based on data from IoV entities while protecting the confidentiality of the data and reducing the risks of tampering with the data. However, ML models created this way are still vulnerable to evasion, poisoning, and exploratory attacks using adversarial examples. This paper investigates the impact of various possible adversarial examples on the BFF-IDS. We proposed integrating a statistical detector to detect and extract unknown adversarial samples. By including the unknown detected samples into the dataset of the detector, we augment the BFF-IDS with an additional model to detect original known attacks and the new adversarial inputs. The statistical adversarial detector confidently detected adversarial examples at the sample size of 50 and 100 input samples. Furthermore, the augmented BFF-IDS (BFF-IDS(AUG)) successfully mitigates the adversarial examples with more than 96% accuracy. With this approach, the model will continue to be augmented in a sandbox whenever an adversarial sample is detected and subsequently adopt the BFF-IDS(AUG) as the active security model. Consequently, the proposed integration of the statistical adversarial detector and the subsequent augmentation of the BFF-IDS with detected adversarial samples provides a sustainable security framework against adversarial examples and other unknown attacks.

## 1. Introduction

With the tremendous development of network communication technology such as 5G, transforming the modern transportation system into the concept of the Internet-of-vehicle (IoV) is becoming a reality [1]. The IoV is envisaged to provide a seamless communication framework for communication between smart vehicles, such as connected vehicles (CV) and autonomous vehicles (AV), and other IoV entities such as pedestrians and road infrastructure/devices [1, 2]. IoV mainly consists of intra-vehicle Networks (IVNs), which consist of an In-vehicle communication system managed by a controller area network (CAN) and an external vehicular network that is concerned with the interaction of vehicles to outer environments by vehicle-to-everything (V2X) technology [2].

The CAN manages the interaction of Electronic Control Units (ECU), which facilitates internal vehicle components such as engine, brake and telemetric systems through information exchange [3]. However, the CAN was designed with no security mechanism to deal with malicious communication broadcast in the CAN bus. This has left it vulnerable to attack through the vehicle's On-Board Diagnostic II (OBD-II) port, firmware (such as media player) or remotely through telematics [4, 5]. Machine Learning (ML) for a CAN Intrusion Detection System (IDS) is a promising solution to protect the CAN due to its ability to learn non-linear attack patterns. Tremendous progress has been made in adopting machine learning for CAN IDS [6-9]. Traditionally, the models are trained locally for a single vehicle due to limited support by automakers and car owners to share sensitive data. Therefore, the models are denied the benefit of access to rich data available in the vehicle ecosystem. The ability to create models based on data from various vehicles in an ecosystem allows for taking into account a higher number and variety of attacks, potentially improving accuracy.

A Blockchain-based Federated Forest Software Defined Networking (SDN)-enabled Intrusion Detection System (BFF-IDS) can be used to support ML that utilizes the data available in the whole ecosystem of vehicles while at the same time protecting sensitive data [10]. In BFF-IDS, each participating vehicle (miner) trains a partial model and stores it at an InterPlanetary File System (IPFS). They exchange the hashes of these partial models via blockchain. The hashes are unique to their models and difficult to change once incorporated into the blockchain. Thus, storing the hashes on the blockchain provides a 'proof of existence' of the partial model [11]. This makes tampering with the models after storing their hashes detectable and thus provides partial protection against poisoning attacks.

Next, the final federated model is obtained by aggregating the partial models at the user end. Tests in a sandbox environment show that BFF-IDS has an accuracy of 98.10% in detecting fuzzy attacks, impersonation attacks, DoS attacks and attack-free traffic. This is considerably higher than other solutions, such as [12-15], which have an accuracy of 92.80%, 97.00%, 97.00% and 98.0%, respectively.

Adversarial ML is a machine learning technique that studies the vulnerabilities of fooling ML models in the face of deceptive input [16]. The deceptive inputs known as adversarial examples can be created by adding imperceptible perturbation to the actual inputs [17]. Adversaries can target the model during training or testing time.

Although the blockchain reduces the risk of model poisoning after its hash was stored and sharing partial models instead of raw data contributes to confidentiality, ML models are often susceptible to adversarial attacks in which inputs are manipulated to cause the model to misclassify [5, 18]. In addition, sophisticated novel attacks are constantly being developed [19]. Providing a sustainable way of keeping up with these attacks is critical for the system's survival.

The adversarial examples pose a tremendous security threat to the adoption of ML as IDS. In particular, the adversarial example transferability has shown that adversarial examples designed to cause one model, $M_1$, to misclassify often cause another model, $M_2$, to misclassify. Thus, this makes it possible to generate adversarial examples in one ML machine and attack another ML system without knowledge or access to the underlying model [18, 20, 21]. Most current studies focus on providing solutions in image datasets, such as generating adversarial examples based on the MNIST dataset and attacking traffic signs in a CV/ AV. Still, there has recently been keen interest in the impact of adversarial examples in other systems, such as security systems [22]. For instance, the impact of adversarial examples against CAN IDS build using Long Short Term Memory (LSTM) was investigated [22]. Fast Gradient Sign Method (FGSM) and Basic Iterative Method (BIM) attacks were explored, and an Adversarial Attack Defending System (AADS) was proposed to counter the attacks

With the use of blockchain, the BFF-IDS offers defence against poisoning but has no measure for evasion attacks which is the most common attack/threat faced by ML models [23]. Besides, the effectiveness of transferability in CAN IDS has primarily remained untested as most studies focused on image classification problems [23]. In practice, more knowledge is needed on its vulnerability against attacks using adversarial examples and how to protect against such attacks to enable deployment of the BFF-IDS on a large scale in practice. Therefore, we set out to obtain the following objectives following the approach of [16] on traditional ML closely:

- Determine the vulnerability of BFF-IDS to adversarial examples.
- Integrate a statistical adversarial detector for the detection of unknown adversarial examples and augment the BFF-IDS to detect the unknown adversarial examples.
- Investigate the robustness of the solutions of the preceding objective.

This study is an extension of the BFF-IDS where we investigated the resilience of BFF-IDS against unknown evasion attacks using the principle of adversarial example transferability to generate adversarial samples. We explore the integration of a statistical (adversarial) detector to check for unknown adversarial attacks. However, statistical attacks can only detect adversarial examples in large batch samples [16]. To protect against single-input attacks, detected attacks are extracted into a sandbox and added to the dataset as an adversarial class to augment the model by training a new BFF-IDS (AUG) model. This way, the proposed system could be made sustainable in detecting unforeseen attacks. In particular, the main contributions of this study are thus summarized as follows:

- We investigated the impact of adversarial examples against BFF-IDS using various

algorithms based on the adversarial sample transferability. Our experimental results suggest that BFF-IDS is very vulnerable to adversarial examples attack as it succeeded in significantly reducing the confidence (accuracy) of our model from more than 97% to as low as 20% in some instances.
- We investigated the integration of a statistical adversarial detector to detect unknown adversarial examples. The integration of the detector effectively detected adversarial samples from benign distributions in large batch samples. However, the statistical test cannot identify which samples are adversarial.
- To address the limitation of the adversarial detector, which can only detect adversarial samples from a group (batch) of samples, we augmented the BFF-IDS by adding the detected samples to the dataset and adding a new class, "adversarial class", to the output of the model. The retrain model significantly improves the model's confidence as attacks can now be detected on single input- bases.
- We demonstrated the robustness of the statistical test by considering the mixture of adversarial samples from various algorithms and benign samples. Also, we investigated the robustness of the BFF-IDS augmentation by investigating how the mixture of several combinations of adversarial examples from the different algorithms as training samples can affect the detection of the adversarial samples and the general performance of the model. Useful results and conclusions were derived.

The rest of the paper is organized as follows: Section 2 presents the study's background, including federated forest, Adversarial Machine learning, sample transferability, and statistical testing. The methodology is given in section 3. Section 4 presents the experimental results on adversarial attacks on BFF-IDS, statistical detection of adversarial examples, BFF-IDS augmentation, and the proposed solutions' robustness. The discussion of results is offered in section 5, while section 6 concludes the paper.

## 2. Background

*2.1. Related work*

IDS for CAN has been proposed to offer protection to CAN using various ML models. For instance, a novel graph-based Gaussian naive Bayes (GGNB) is proposed for the intrusion detection algorithm. The GGNB leverages graph properties to detect CAN-monitoring attacks without protocol modification [24]. Although the method recorded promising results, the authors did not investigate the model's impact on adversarial example attacks. Meanwhile, a complex value neural network (CVNN) has been proposed to protect the CAN network to detect the arbitration field [25]. Encoders were employed to extract valuable features for better generalization. However, this study did not consider adversarial examples and data sharing problem

With the innovation of CAVs, CAN and FlexRay are being replaced with Automotive Ethernet to support high-definition applications demand for high throughput etc. Jeong, et al. [26] offered the first intrusion detection method in this domain to detect audio-video transport protocol (AVTP) stream injection attacks using feature generation and convolutional neural network (CNN). Although the method proves the approach is suitable for real-time detection, adversarial examples impact and possible countermeasure were not considered.

Although previous studies on adversarial examples focused on the image domain, considering the impact of adversarial examples on IDS is gaining traction [22]. For instance, Yang et al. [23] demonstrate the vulnerability of deep neural networks for NIDS to adversarial examples using model substitution and black-box-based zeroth-order optimization (ZOO) and generative adversarial network (GAN) attacks. In another study, a testbed consisting of an adversarial model embedded in the IDS was designed to facilitate security evaluation of CAN system design [27]. However, the developed tool provides no function for testing mitigation measures.

Furthermore, adversarial examples have been investigated in connected and autonomous vehicles (CAV) [28]. Although valuable results have been observed on the impact of adversarial examples on several ML and deep learning models, the paper failed to provide countermeasures against threats. Additionally, adversarial examples in spam filters, biometric authentication and fraud detection have been studied in [29-33]. A recent study investigated the impact of adversarial examples against CAN IDS build using Long Short Term Memory (LSTM) [22]. Fast Gradient Sign Method (FGSM) and Basic Iterative Method (BIM) attacks were explored and had a success rate of about 98%. An Adversarial Attack Defending System (AADS) was developed to counter the attacks by retraining the LSTM model with the attack samples as part of the training data.

As illustrated in Table 1, most existing studies are focused on traditional training methods. Besides, adversarial examples' impact on the proposed IDS is mainly unexplored. Therefore, this paper focuses on adversarial examples' impact and countermeasure on BFF-IDS.

**Table 1**
Related works

| Year/Ref. | IDS Domain | IDS Model | BC | FL | SDN | Adv. Ex. impact | Adv. Ex. mitigation |
|---|---|---|---|---|---|---|---|
| 2021, [22] | CAN | LSTM | ✓ | ✗ | ✗ | ✓ (intra-technique) | ✓ |
| 2021, [26] | Automotive Ethernet | CNN | ✗ | ✗ | ✗ | ✗ | ✗ |
| 2021, [24] | CAN | Graph-based Gaussian naive Bayes (GGNB) | ✗ | ✗ | ✗ | ✗ | ✗ |
| 2021, [25] | CAN | CVNN | ✗ | ✗ | ✗ | ✗ | ✗ |
| 2021, [10] | CAN | BFF-IDS | ✓ | ✓ | ✓ | ✗ | ✗ |
| 2019, [27] | CAN | K-Nearest Neighbor (k-NN) | ✗ | ✗ | ✗ | ✗ | ✗ |
| 2019, [28] | CAV | k-NN, Random forest (RF), Logistic regression (LR), Long Short-Term Memory (LSTM) | ✗ | ✗ | ✗ | ✓ | ✗ |
| 2018, [23] | NIDS | Deep Neural Network | ✗ | ✗ | ✗ | ✓ (intra-technique) | ✗ |
| Our framework | CAN | BFF-IDS | ✓ | ✓ | ✓ | ✓ (cross-technique) | ✓ |

*Adv Ex: adversary examples, BC: Blockchain, FL: Federated Learning*

## 2.2. Federated Forest

Federated learning (FL) is an innovative concept in which models are built on data sets distributed across multiple devices while preventing data leakage [34]. Conventionally, a model, $\mathcal{M}_{SUM}$, is trained by $\{\mathbb{N}_1 \ldots \mathbb{N}_i\}$, who wish to build a stronger model by combining data, $(\mathcal{D}_1 \ldots \mathcal{D}_i)$, from data owners, consolidating their data, $\mathcal{D} = \mathcal{D}_1 \cup \ldots \cup \mathcal{D}_i$. Unlike the traditional learning (TL) method, FL enables the collaborative training of the model, $\mathcal{M}_{FL}$, in such a way that the confidentiality of the data $\mathcal{D}_i$ from any owner $\mathbb{N}_i$ is preserved while ensuring the performance, $\rho_{FL}$, of the model, $\mathcal{M}_{FL}$, is close to the performance, $\rho_{SUM}$, of the model, $\mathcal{M}_{SUM}$. Formally, given $\varphi$ as a non-negative real number, if

$$|\rho_{FL} - \rho_{SUM}| < \varphi \quad (1)$$

the federated model is said to have $\varphi$-accuracy loss.

Based on the data distribution among owners (subsets), FL can be categorized into horizontal FL, vertical FL and federated transfer learning [34]. For horizontal FL, the data set in each subset have the same feature space but a different number of samples. The datasets have the same number of samples in vertical FL but different feature spaces. A scenario where both the sample and feature space differ is designated as federated transfer learning.

For this study, horizontal FL is utilized to build federated forest, $i$, for intrusion detection. Given the dataset $\mathcal{D}$ distributed among $k$ federating units (owners), $\mathbb{N}_1 \ldots \mathbb{N}_k$ are assumed to be disjoint such that only a subset of the data $\mathcal{D}_k \subseteq \mathcal{D}$ with $N_k$ samples are used by $k^{th}$ unit, where $k \in [1,k]$. The goal is to build an accurate federated forest model such that: (1) a partial random forest model is trained and held by each owner (known as a miner), $\mathcal{M}_i, 1 \leq i \leq k$; (2) the FL model, $\mathcal{M}_{FL}$, is aggregated at each user end while minimizing $\varphi$-accuracy loss.

## 2.3. Adversarial Machine learning and samples transferability

Formally, let's assume the ML model, $M$, correctly classified benign sample S, i.e. $M(S) = y_{true}$. An adversarial example, $A$, can be constructed such that it is perceptually indistinguishable from $S$ but causes the model to misclassify. i.e. $M(A) \neq y_{true}$.

The adversarial example is crafted by adding a small perturbation, $\delta$, to the benign sample $S$. The $\delta$ is computed by the approximation of the following optimization problem iteratively until it gets classified by the by ML classifier:

$$A = S + \delta_s \quad (2)$$

where $\delta_s = \arg\min_{\delta} M(S + \delta) \neq M(S)$

Adversarial examples transferability refers to the potential of adversarial examples generated and design for model $M$ to also cause the misclassification in $M'$ without access to the underlying model [18].

Considering crafting the adversarial examples by solving the optimization problem, we can formalize the adversarial sample transferability notion as:

$$\Omega_X(M, M') = |\{M'(S) \neq M'(S + \delta_s) : S \in X\}| \quad (3)$$

where set $X$ is the expected task input distribution solved by models $M$ and $M'$.

The adversarial sample transferability can be categorized into intra-technique and cross-technique [20]. In intra-technique transferability, the cross models are trained with the same ML technique but different parameter initializations or datasets- For instance, both $M$ and $M'$ are neural networks or decision trees. The cross-technique transferability deals with a situation where the models are trained using different techniques-e.g., $M$ is a neural network and $M'$ is a decision tree. In this study we focus on cross-technique transferability as it represents the real-life scenario of how attacks are conducted.

*2.4. Statistical Hypothesis Testing*

Two-sample hypothetical testing is introduced to conduct a test on two randomly selected samples to determine the statistically significant difference between the two samples, in other words, whether the samples originated from the same distribution. Credit for the statistical hypothesis and analysis goes to [16].

Formally, let $S \sim p$, notation be considered as a sample drawn from distribution $P$. A statistical test can be formalized as follows: Given $S_1 \sim p$ and $S_2 \sim q$ where $S_1 = n$ and $S_2 = m$; the null hypothesis $H_0$ holds that $p = q$. The statistical test $\mathcal{T}(S_1, S_2): S^n \times S^m \to \{0,1\}$ takes in the sample's input and returns the p-value, which matches the significant level, $\alpha$. The p-value gives the probability of obtaining the observed outcome or a more extreme one, while the $\alpha$ relates to the confidence of the test set as the threshold. In this study, we set the threshold at 0.05. Therefore, the $H_0$ is rejected if the p-value is less than the threshold.

Several two-sample tests have been proposed, but we adopt kernel-based, which measures the probability of the distance between the two samples as a biased estimator of the true Maximum Mean Discrepancy (MMD). We also compared the MMD with energy distance (ED) [16].

*2.5 CAN Bus Dataset*

The CAN-intrusion dataset (OTIDS) used in this study was obtained from the Hacking and Countermeasure Research Lab at Korea University [28]. The dataset was obtained from real attack scenarios and consists of four classes of traffic: fuzzy attack, DoS attack, Impersonation attack, and attack-free state. The datasets were created from a real vehicle (KIA SOUL) by logging onto CAN traffic via the OBD-II port. These attack types have devastating consequences on the CAN- the fuzzy attack can override normal function; a DoS attack can deny access to a legitimate node, while an impersonating attack can cause the vehicle to manifest an unintended state or action.

## 3. Methodology

*3.1. BFF-IDS for IoV security*

With the advent of smart cities and IoV, the network architecture is increasingly faced with high-performance demand regarding latency, scalability, network bandwidth usage, data privacy and security [35]. The problem is exacerbated by the diverse technologies and a high degree of interdependence between various system components in the network ecosystem [36]. This study proposed a hybrid architecture that will guarantee network scalability and privacy using blockchain and SDN. This architecture was initially proposed by Kim, et al. [35], but their system is expensive in terms of gas/ether (as in Ethereum) as model training parameters are exchanged through the blockchain.

The hybrid architecture for the IDS consists of three planes-the data plane consisting of the individual vehicle CAN; the control plane, which manages the interaction of the CANs through blockchain and the control plane, which entails system management authorities(stakeholders). Each vehicle (known as a user node) builts a partial model using its data and upload the model to IPFS while the location hash is exchanged through the blockchain- unlike the approach of [37], only the cost of some bytes is incurred in the process. The blockchain network consists of miner nodes responsible for creating blocks and verifying proof-of-Authority. The vehicles are SDN enabled to aid ease deployment costs with high agility and security. The stakeholders include network management, ID providers and threat intelligence agencies that further investigate attack trends and other security policies.

To better expose the system requirement considering the complex interaction of the system component-SDN, Ethereum Blockchain, IPFS and machine learning libraries- object transformation and event effects on the system's behaviour must be examined to build a testbed. We adopt structured analysis to present the requirements modeling in which data and transformation processes are treated as separate entities [38]. Behavioral modeling is considered here as it effectively exposes the testbed design's requirements for the structured requirement analysis.

The dynamic behaviour modeling of the testbed during operation is accomplished by representing the various testbed components processes as a function of events and time. It exposed the details of the testbed response to external events (i.e., system components). We utilized a sequence diagram (SD) to illustrate how events caused transitions between components. Fig. 1 presents a sequence diagram for the testbed operation, illustrating events and corresponding transitions between components. The arrows depict the event-driven transition/behaviour between components. The time of event occurrence is measured vertically downward along each component. The vertical rectangles along the components represent the time spent processing and the activity. Creating network topology, partial model training, and aggregation takes more time in each corresponding component of the testbed. At the end of the model aggregation, the session of FL miner and users are terminated by the Mininet emulator. More details on the testbed in [10, 39].

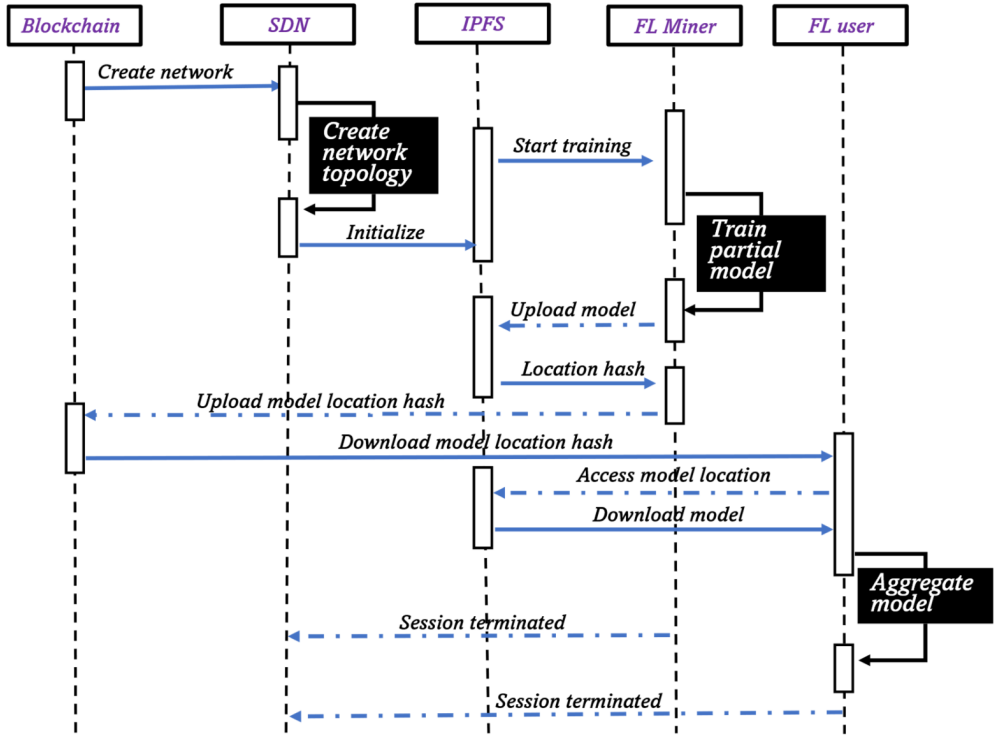

**Fig. 1.** Testbed sequence diagram

### 3.2. Threat Model

This section identifies the fundamental security objectives, threats, and vulnerabilities of the BFF-IDS, considering the adversary's strength, goals, knowledge, and capabilities. Firstly, we highlight the attack surface in the BFF-IDS through which the adversary may attempt to launch an attack to subvert the system. We then decompose the threat model into four aspects: adversarial knowledge, capabilities, specificity and goals.

***The Attack Surface.*** This study defined the attack surface concerning the data processing pipeline, including the injection and feature extraction stages. We assume the adversary launched the attack to corrupt the BFF-IDS based on its knowledge of the feature extraction and access to source traffic. There are three possible attack scenarios at the attack surface: evasion, poisoning, and exploratory attack [40]. The Evasion attack is possible during the testing phase, whereby the adversary manipulates test data to corrupt the model. The poisoning attack occurred during the training phase, in which the training data is contaminated to compromise the whole learning process, while exploratory uses the black-box approach to learn about the underlying model and the training data pattern. In this study, we investigate the evasion attack on the BFF-IDS by manipulating the extracted feature (see Fig. 2 for the illustration of the threat model).

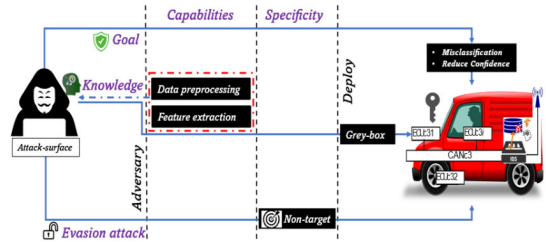

**Fig. 2.** Threat Model

***Adversarial Knowledge.*** Adversarial knowledge can be divided into three categories: white box, black box and grey box [17]. In a white-box attack, the adversary is assumed to have complete knowledge of the underlying model and the dataset. In the block-box attack, the adversary does not know the underlying model and access to the training data, whereas, in the Gray-box attacks, the adversary is assumed to have partial knowledge of the target model. Since we are investigating cross-technique transferability, we crafted the adversarial examples (features) using various models based on white-box attacks. The actual attack on the BFF-IDS is conducted in a block-box, assuming the adversary only knows the feature extraction. Thus, we categorize our approach to be a grey-box attack.

*Adversarial Capabilities.* In this study, the adversary is assumed to be able to manipulate features extracted during the testing/deploying phase to corrupt the BFF-IDS. Neither the trained model, process, nor data is affected. Specifically, the adversary can draw surrogate samples from the original distribution of the benign dataset. But the attacker does not know the model algorithm or design. Thus, the attack is conducted on the trained BFF-IDS using the crafted surrogate feature at the testing/deployment phase.

*Adversarial Specificity.* The adversarial examples can be crafted to compromise the model's performance on a specific class (targeted attack) or to reduce the model's classification performance confidence (non-targeted attack). However, this study is limited to non-target attacks.

*Adversarial Goals:* In this study, the adversary's goal includes confidence reduction and misclassification. The adversary aims to reduce the confidence of the BFF-IDS by causing output ambiguity and increasing misclassification by altering the detection output class to any class different from the original class.

*3.3. Adversarial examples/unknown attack detection Framework for BFF-IDS*

In the framework (see Fig. 3), the BFF-IDS Model is built using federated learning in which each miner partially trains its model and uploads it into IPFS, while the pointers to the model are stored in blockchain. Legitimate users can then download the models through the hash obtained from the blockchain. The federated model is then finally aggregated at the users' end. The detailed design and implementation of the BFF-IDS are provided in [10]-the CAN ID cycle (frequency of occurrence) is extracted and transformed using Fast Fourier Transform(FFT); statistical and entropy features are then extracted. The features include minimum, maximum, mean, standard deviation, skewness, kurtosis, Shannon entropy, sample entropy and permutation entropy. The attack is assumed to be launched after deployment in black-box scenario-the adversary does not know the BFF-IDS underlying structure. We investigate the effect of various adversarial attack methods on the BFF-IDS, which include the Fast Gradient Sign Method (FGSMA), Jacobian-based Saliency Map Approach (JSMA), Support Vector Machine (SVM) and Decision Tree (DT) attacks. Motivated by [16], we proposed the integration of a statistical adversarial detector before the BFF-IDS. The detector utilized a statistical method (MMD, ED) to detect the adversaries and unknown attack patterns. Detected attack traffic is then captured and saved into a sandbox to retrain new BFF-IDS with an "adversarial attack" class as a new class of the detected traffic. Therefore, each user is expected to have the detector in its CAN system to monitor unknown/adversarial attacks and retrain its model using its old datasets and the new adversarial data. The adversarial attack class would be used as samples to train the model. This way, the BFF-IDS can be sustainable as a new version would release over time to withstand any new adversarial or novel attack patterns.

*3.4. Adversarial examples crafting*

This section presents the different adversarial generating methods employed in this work. Although there is no guarantee that the method described here will generate traffic that BFF-IDS will misclassify, the degenerated samples are described as "adversarial examples" or "unknown attacks". We employed differentiable machine learning models such as DNN and non-differentiable machine learning models such as Support Vector Machine (SVM) and Decision Tree (DT) to craft adversarial samples.

*FGSMA.* FGSM attack was proposed by Goodfellow [41]- This method utilizes the gradient of a model's output to perturb adversarial examples, $\varkappa'$, with respect to its input in that direction. Although the method is computational efficient, it introduces a significant perturbation that distorts the input distribution-This may not be acceptable in some domains [23]. The perturbation can be defined as:

$$\eta = \epsilon sign(\nabla_x J_\theta(x,l)) \qquad (4)$$

where $\epsilon$ is the magnitude.

*JSMA.* Papenot proposed JSMA to address the problem of FGSMA by reducing the scale of perturbation through the iterative computation of the best feature to perturb for misclassification [42]. This approach enables the extraction of the influence of an individual feature on a particular class through a saliency map. In contrast, other features introduce perturbation on the original input, resulting in a misclassification [23]. However, the computational cost of JSMA is much significant. The JSMA attacks utilize the Jacobian matrix to evaluate the model's output sensitivity. The matrix is expressed as:

$$J_f(x) = \frac{\partial f(x)}{\partial y} = \left[\frac{\partial f_j(x)}{\partial x_i}\right]_{i \times j} \qquad (5)$$

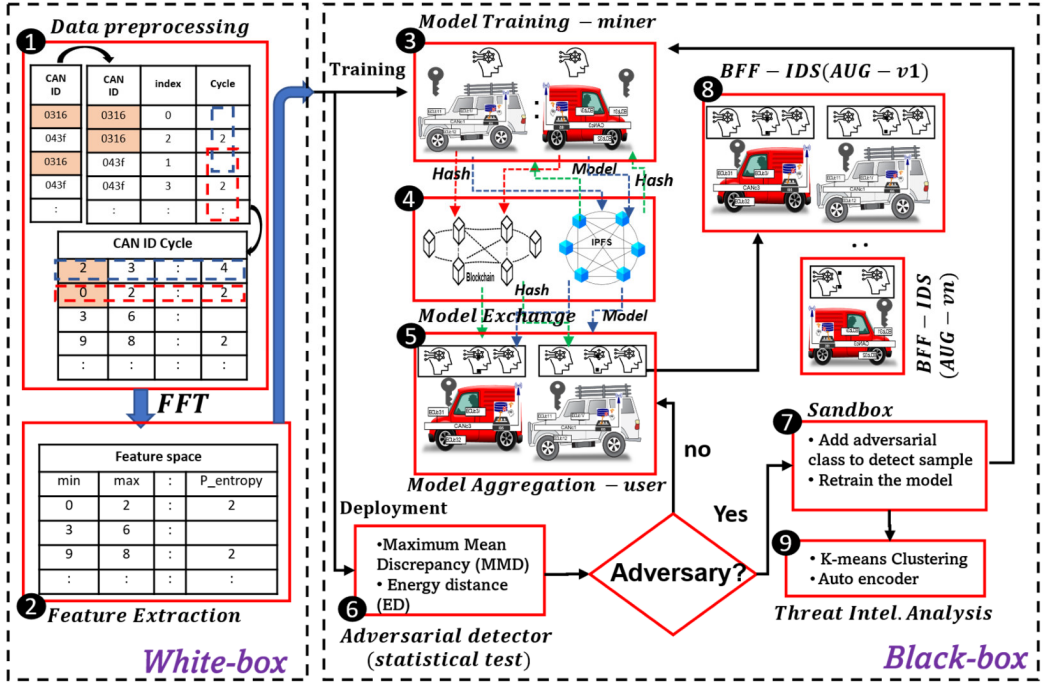

**Fig. 3.** Adversarial Detection Framework Overview

*SVM Attack.* This method attacks SVM by selecting a point orthogonally in the direction of the hyperplane, acting as the decision boundary to the SVM subclassifier [20]. The points are selected using

$$\Phi_{\rho+1} = \left\{ \vec{x} - \lambda \cdot \frac{\vec{\omega}\left[\tilde{O}(\vec{x})\right]}{\left\|\vec{\omega}\left[\tilde{O}(\vec{x})\right]\right\|} \vec{x} : \vec{x} \epsilon \Phi_\rho \right\} \cup \Phi_\rho \quad (6)$$

where $\Phi_\rho$, $\Phi_{\rho+1}$ and $\lambda$ are the previous, new training sets and a fine-tuning parameter for augmentation step size, respectively; $\vec{\omega}[k]$ is the weight that represents the hyperplane direction of subclassifier $k$ used for the implementation of a multi-class SVM.

*DT attack.* In the DT attack, the shortest path is computed between the current leaf at which the sample is and the nearest leaf of another class. The feature in the first common node shared by the two paths is repeatedly perturbed until misclassification is achieved by modifying a few non-targeted features [16].

### 3.5. Statistical Metrics for Adversarial examples/unknown attacks detection and BFF-IDS Augmentation

#### 3.5.1. Statistical Metrics for Adversarial examples/unknown attacks detection

To distinguish between a known/benign sample distribution and an adversarial example (unknown samples) distribution, we investigate two statistical distance measures: MMD and ED. Motivated by the hypothesis of [16], the first hypothesis is to determine whether MMD and ED can distinguish between the benign CAN feature and adversarial examples generated from them.

*"Hypothesis 1.* Measurable difference between known benign samples and adversarial examples can be observed within a bounded number of *n* examples using a consistent statistical test *T*."

However, this hypothesis is limited to (1) the finite number of samples needed to observe the difference and (2) the unknown attack detection is restricted to the adversarial examples crafting algorithm. The validation of this hypothesis is presented in section 4.

Formally, the statistical divergence measures are defined as:

*Maximum Mean Discrepancy (MMD)*—Given two random samples, $X_1$ (benign) and $X_2$ (adversarial), the MMD, which measures the distance/divergence between the two samples, is formalized as:

$$MMD_b[\mathcal{F}, X_1, X_2]$$
$$= \sup_{f \epsilon \mathcal{F}} \left( \frac{1}{n} \sum_{i=1}^{n} f(x_{1i}) - \frac{1}{m} \sum_{i=1}^{m} f(x_{2i}) \right) \quad (7)$$

where $x_{1i} \in X_1$, $x_{2i} \in X_2$ are the *i*-th data point in the first and second samples, respectively. Kernel function $f \epsilon \mathcal{F}$ is selected to maximize the distances between the samples. In our case, we use a Gaussian kernel.

*Energy distance (ED)*—we also employed the ED to measure the statistical distance between benign data and adversarial examples distributions. The ED is a specific case of MMD in which no kernel is applied [16]. Formally, for a *d*-dimensional random sample, $X_1$(benign) and $X_2$ (adversarial), the ED is defined as [43]:

$$\varepsilon(X_1, X_2) = 2E|X_1 - X_2|_d - E|X_1 - X_1'|_d - 2E|X_2 - X_2'|_d \quad (8)$$

where $E|X_1| < \infty$ $E|X_2| < \infty$, $X_1'$ is an independent and identically distributed (iid) copy of the $X_1$, and $X_2'$ is an iid copy of $X_2$.

The statistical test would be beneficial in monitoring and detecting adversarial examples and unknown traffic in the CAN. Each user node can collect any unknown detected samples for further analysis and retrain the model to detect the new unknown attacks. However, the statistical test only detects adversarial/unknown samples in large batches [16]. This means that a single input attack will go undetected, thus the statistical test is not suitable for offering security for the CAN.

*3.5.2. BFF-IDS Augmentation*

To address the problem of detecting single attack samples, the detected adversarial/unknown attacks in large batches by the statistical test can be utilized to retrain the model by augmenting the BFF-IDS with an additional "unknown" class $Y_{adv}$. The new model can, after that, replace the old version. With this approach, regular updates or model versions can be retrained and released whenever a new attack trend is detected. Therefore, motivated by [16], the hypothesis is on whether augmenting the BFF-IDS model to detect new attacks is suitable for CAN security:

*"Hypothesis 2.* Augmenting the BFF-IDS model with unknown class training samples can successfully detect adversarial examples and other unknown samples."

The goal of the BFF-IDS augmentation is for the model to be able to detect single adversarial attacks. In the augmentation of the BFF-IDS, the initial original test dataset, $D_{real} = \{X, Y\}$, where X and Y are features and label is used to craft adversarial features, $X_{adv}$. Different crafting algorithms are employed to generate the samples, which are then assigned to one class, $Y_{adv}$. A new model, BFF-IDS (AUG), is trained on the augmented dataset $X \cup X_{adv}$ with the $X$ having Y original label and all adversarial samples belonging to the same class, $Y_{adv}$.

In an actual application, these $X_{adv}$ or unknown attack distribution is detected by the statistical test. Then existing datasets are augmented with the newly detected class, and the model is retrained. The statistical detectors are then updated with the augmented samples as the new known samples for testing incoming traffic. The validation of this hypothesis is presented in section 4.

*3.6. Performance Evaluation Metric*

The model performances are investigated using precision, recall, F1-score, and accuracy. The investigation includes the model's performance on benign data, adversarial examples, and the BFF-IDS performance after augmentation. In addition, the robustness of the statistical test detector and the model augmentation are also investigated. The evaluation metrics are expressed as follows:

$$Accuracy = \frac{TP + TN}{TP + TN + FP + FN} \quad (9)$$

$$Precision = \frac{TP}{TP + FP} \quad (10)$$

$$Recall = \frac{TP}{TP + FN} \quad (11)$$

$$F_1 = \frac{2TP}{2TP + FP + FN} \quad (12)$$

where TP, TN, FP, and FN are the number of true-positive, true-negative, false-positive, and false-negative cases, respectively.

**4. Experimental Results**

This section presents the experimental results of our investigation of various attack models on BFF-IDS. The detailed implementation, training and test results for the BFF-IDS are presented in [10]. Therefore, at this point, we assume the federated model is trained and deployed. We first present the adversarial examples crafted and the statistical test result of the adversarial samples with benign samples. We then present the detection result of benign data and the adversarial samples by various users' nodes in the BFF-IDS model. Furthermore, we present the results regarding adversarial detectors and model augmentation.

Firstly, we investigated the performance of the BFF-IDS on benign test data. The BFF-IDS are built using miners ranging from 5 to 20. As indicated in Fig. 4, 5 miners' model has the best generalization with the score of 0.97611, 0.97516 and 0.97540 for precision, recall and F1- score, respectively. The lowest-performing model is that of 20 miners with records of 0.94345, 0.93936 and 0.94027 for precision, recall and F1-score, respectively. As evident from the results, the model's performance decreases with an increase in the number of miners. This is a result of the splitting of the training data based on the number of miners. The higher the number of miners, the fewer the number of the dataset available for training each model. Therefore, the 5 miners' model has enough data for training and thus better generalization result of benign test data.

*4.1. Identification of Adversarial examples using statistical metrics and test*

This section presents the statistical divergence measure on both benign and adversarial datasets using the MMD and ED. For the FGSM attack, we varied the perturbation from 0.01 to 0.50 to observe how the statistical score varies. As indicated in Table 2, the MMD and ED of adversarial samples generated by FGSM increased with perturbation $\epsilon$. Except for JSMA, the MMD values of adversarial samples are higher than that of benign data. Likewise, there is an increase in the ED values for the adversarial samples except for the DT attack. Both JSMA and DT attacks show little increase in the MMD and ED, respectively, compared to the benign data. Fig. 5 depicts the distribution of the statistical measures across the benign and adversarial examples, with the MMD and ED ranging from 0.01856 to 0. 4027 and 2.8576 to 4.1742, respectively. Consequently, these results suggest a considerable statistical distinction between benign samples and generated adversarial samples. Thus, the statistical approach is sufficient to detect adversarial attacks against BFF-IDS.

*4.2 Adversarial examples against BFF-IDS*

In this section, we attack the BFF-IDS using the generated adversarial examples. The use of various adversarial crafting models to attack BFF-IDS is based on the principle of cross-technique adversarial sample transferability [20]. This principle is considered in our investigation because an adversary can launch any form of attack, which might be different from what the BFF-IDS was trained to detect- The adversary can generate attack traffic on the CAN using an adversarial machine or model.

The investigation focuses on the best model, i.e. 5 miners, BFF-IDS. Fig. 6 presents the performance of the BFF-IDS under attack. The performance of the model significantly drops from the score of 0.97611, 0.97516, and 0.9754 in precision, recall and F1-score to about 0.12064 (SVM attack), 0.17828 (FGSM 0.36) and 0.13523(SVM attack), respectively. The DT attack recorded the second least degradation in performance of about 0.3 to 0.4 across the metric. On the other hand, the JSMA attack was unsuccessful in degrading the performance of the BFF-IDS. As shown in Table 2, the MDD value for the JSMA data was lower than the benign data.

Furthermore, we investigate the Federated Learning situation where each model, BFF-IDSi, is aggregated at the user end, i. We, therefore, assume that the attack is conducted at the user end of various numbers ranging from 5 to 40. The benign and adversarial samples are split equally based on the number of users

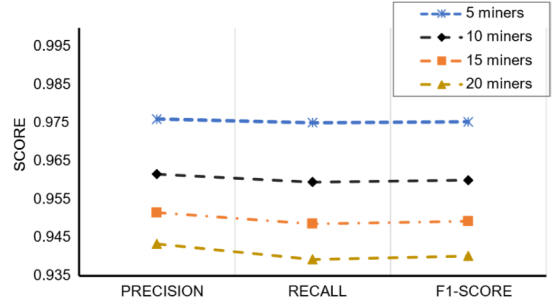

**Fig. 4.** BFF-IDS Performance on Benign Test Data

Table 2:

Maximum Mean Discrepancy (MMD) and Energy Distance (ED) between the Benign Distribution and Adversarial Samples

| Manipulation | $\epsilon$ | MMD | ED |
|---|---|---|---|
| Benign | | 0.0226 | 2.8654 |
| FGSM | 0.01 | 0.0278 | 2.8707 |
| FGSM | 0.08 | 0.0422 | 2.9251 |
| FGSM | 0.15 | 0.0855 | 3.0425 |
| FGSM | 0.22 | 0.1371 | 3.2086 |
| FGSM | 0.29 | 0.1953 | 3.4128 |
| FGSM | 0.36 | 0.2522 | 3.6306 |
| FGSM | 0.40 | 0.2917 | 3.7789 |
| FGSM | 0.50 | 0.4027 | 4.1742 |
| JSMA | | 0.0185 | 2.8698 |
| DT attack | | 0.0599 | 2.8576 |
| SVM attack | | 0.252 | 3.3947 |

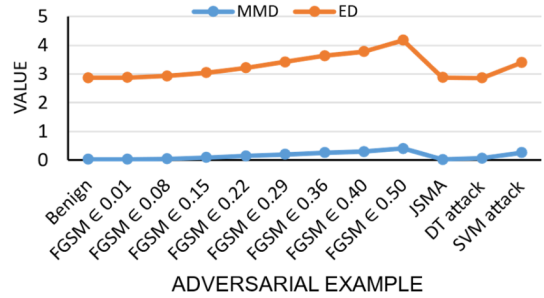

**Fig. 5.** Statistical Measures for Benign and Adversarial Examples Distribution

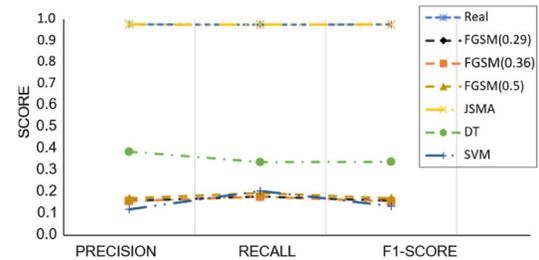

**Fig. 6.** BFF-IDS Performance on Benign Data and Adversarial Samples

under consideration. As indicated in the results as shown in Fig. 7, except for JSMA, the adversarial samples succeeded in significantly degrading accuracy from about 0.98 to the least accuracy of about 0.18,0.18, 0.20, 0.20 and 0.34 for FGSM (eps 0.29), FGSM (eps 0.29), FGSM (eps 0.29), SVM and DT, respectively. The least performance for all cases was recorded in 5 users' scenarios.

*4.3 Statistical Hypothesis for Adversarial examples detection*

Based on the hypothesis, $H_0$, that "benign data are statistically closed to the training data distribution", we compute the MMD (using Gaussian Kernel) and its corresponding p-values as implemented by [16]. The p-values are then compared against the threshold set to 0.05. For legitimate samples, the p-value is expected to be higher than the threshold for the hypothesis to hold. The sample size is critical in detecting the difference between benign and adversarial samples. It becomes more complex as we deal with a dataset containing four classes of attack (DOS, fuzzy, impersonation, and attack-free).

We conducted several experiments with sample sizes of 50, 100, 500 and 1000 to obtain the minimum sample size required to detect adversarial distribution for each class of attack. As indicated in Table 2, 50 samples size is sufficient for most cases to discern the adversarial sample from the benign sample and reject the $H_0$-. This is remarkable when compared to the training size of about 80,000 samples.

However, about 100 and 500 sample sizes are needed for impersonation and fuzzy attack to be detected in FGSM ($\in$ 0.36) and JSMA generated samples, respectively. For DT and JSMA-generated impersonation attacks, the statistical test fails to detect the adversarial samples from the benign sample. For lower perturbation $\in$ of 0.29 and less in FGSM, the statistical test could not detect the adversarial samples. Likewise, the statistical test fails to detect the adversarial samples for DT-generated adversarial samples, except for DOS attack samples.

These results are consistent with the result in 4.1, which shows that lower $\in$ in FGSM, DT and JSMA yielded adversarial samples that are less distinguishable from benign samples. For FGSM with high $\in$ (greater than 0.36) and SVM, which recorded high values in both MMD and ED compared to benign samples, were easily detected by the two-sample statistical test-50 samples size was sufficient to reject the $H_0$.

Furthermore, we further investigated the acceptance of $H_0$ based on the adversarial generation method containing a random collection of all attack classes. This approach aims to see the minimum size required when the class-wise approach is not applied-i.e. the adversary launch attack using randomly generated adversarial samples for all the attack classes.

**Table 3.**

Minimum Samples (Adversarial Examples) Size Required to Detect Adversarial Examples Confidently

| Manipul. | Attack Class | | | |
| --- | --- | --- | --- | --- |
| | DOS | Fuzzy | Impers. | Attack-free |
| FGSM ($\in$ 0.36) | 50 | 50 | 100 | 50 |
| FGSM ($\in$ 0.50) | 50 | 50 | 50 | 50 |
| JSMA | 50 | 500 | - | 50 |
| SVM | 50 | 50 | 50 | 50 |
| DT | 500 | - | - | - |

As presented in Fig. 8, only FGSM ($\in$ 0.5) was detected with a sample size of 50. Meanwhile, FGSM ($\in$ 0.36) and SVM-generated adversarial samples were detected at a sample size of 100. However, the statistical test also fails to detect JSMA and DT-generated adversarial distribution in this case, confirming the earlier results in section 4.1. Thus, it requires more sample size to detect adversarial sample distribution containing random classes of all the attacks than class-wise statistical tests on such a similar distribution.

*4.4 BFF-IDS Augmentation for Adversarial examples mitigation*

The previous section observed that the adversarial sample distribution differs statistically from the benign sample distribution. However, the statistical test cannot detect adversarial samples on a single-input basis and its confidence diminishes with the decrease in the number of samples in a batch. More so, the statistical test cannot pinpoint which input is adversarial in a group of sample-this is consistent with the findings of [16].

In this section, experimental results regarding hypothesis 2 are provided. The augmentation BFF-IDS model by training the model with the addition of samples having adversarial class as labels should be effective in detecting attacks in the benign data and adversarial samples.

We conducted two experiments; in the first experiment, we considered FGSMA (0.36) and SVM as the adversarial samples to be augmented into the

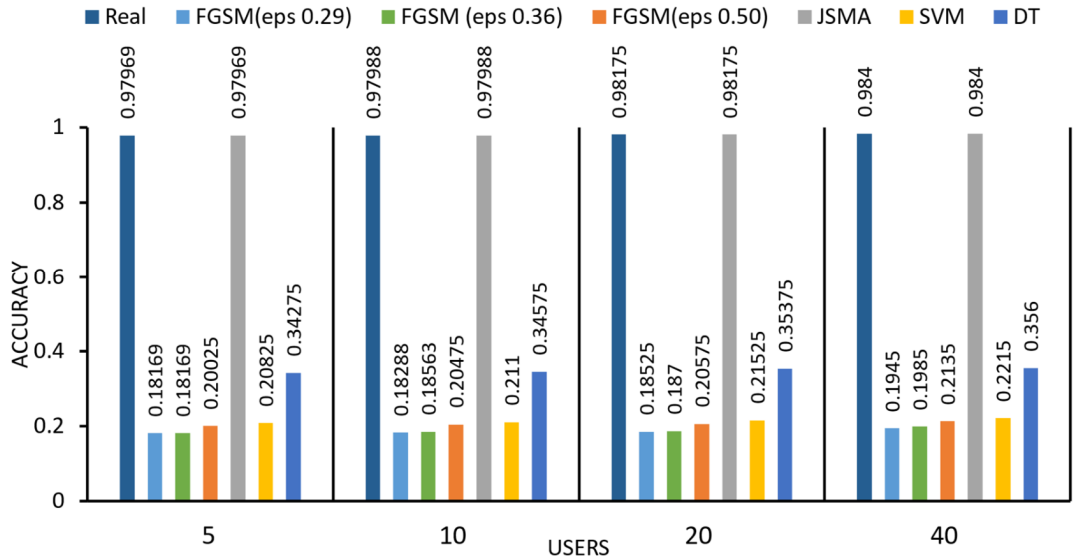
**Fig. 7.** BFF-IDS Performance for Different users on Benign Data and Adversarial Samples

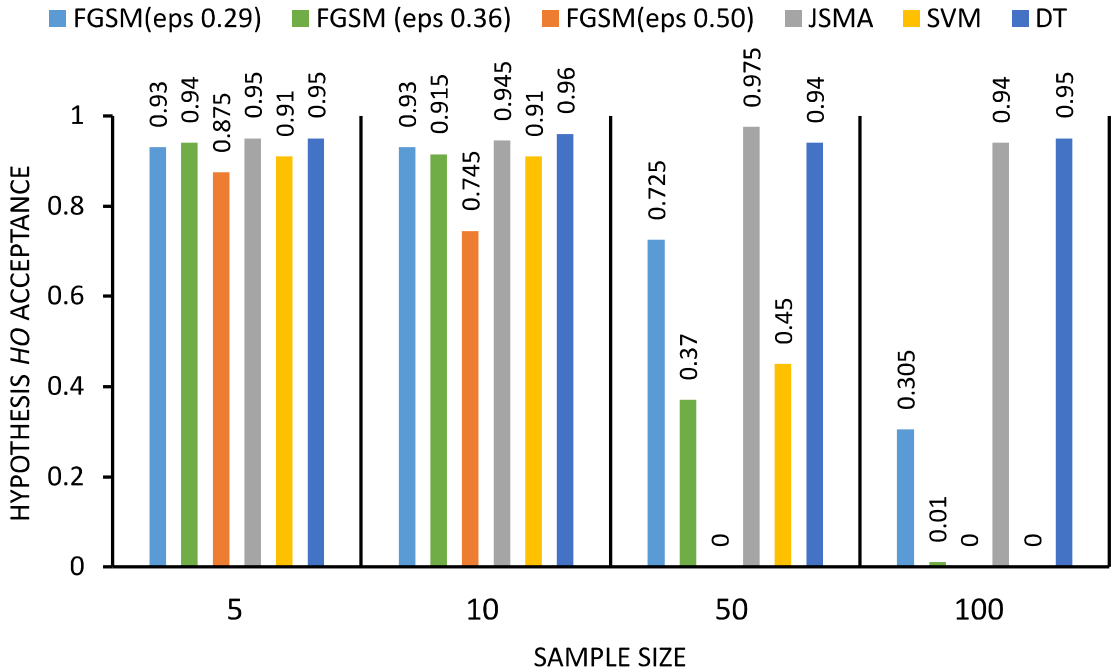
**Fig. 8.** Hypothesis $H_o$ Acceptance concerning the Sample Size based on the Statistical Test

dataset as an adversarial class. These adversarial samples were considered because the statistical test results in the previous section show that these samples are easily detected with fewer samples. The new trained augmented model, BFF-IDS(AUG), is then tested on the adversarial samples, including those not included in the training, as presented in Table 4. The detection accuracy among FGSM and SVM- attacks increased to more than 76.20 % and 79.20%, respectively. The improvement results from including these samples in training as part of the adversarial class. On the other hand, the accuracy of JSMA and DT-attack reduced to -97.46% and -33.17%, respectively. The reduction results from the exclusion of these samples in the training set. Furthermore, the FGSM set for all the value $\in$ detection saw improvement despite only $\in$ 0.36 is used for the training.

The second experiment considered all the adversarial generation algorithms, i.e., FGSMA (0.36), SVM, JSMA and DT, as training samples belonging to the adversarial class. As presented in Table 5, except for JSMA, the detection rate of all the adversarial samples improved, including that of DT, which shows significantly low MMD values compared to others. The DT detection rate increased to about 56.17%.

Considering that BFF-IDS is a federated model, we consider the best performances of the 5-miner model across various numbers of users. For the first experiment with only FGSMA (0.36) and SVM as augmented as training samples, the detection rates for the FGSMA and SVM across all users were more than 0.96, as shown in Fig. 9. As expected, the detection rate for the JSMA and DT were low across all the users and were between 0.001 to 0.009. Also, the detection rate for the second experiment with all the adversaries included in the training increased across all users except for JSMA, which shows little or no improvement, as shown in Fig. 10. This is also consistent with the statistical test finding in the previous section. Therefore, augmenting the model with an adversarial class would improve the detection rate of the model in a single-input adversarial example attack.

**Table 4.**
BFF-IDS Augmentation using FGSM 0.36 and SVM

| Manipul. | BFF-IDS Detect. Rate | BFF-IDS (AUG) Detect. Rate | Recovered rate by the BFF-IDS (AUG) |
|---|---|---|---|
| FGSM (∈ 0.29) | 0.17919 | 0.96140 | 78.221 % |
| FGSM (∈ 0.36) | 0.17860 | 0.96345 | 78.485% |
| FGSM (∈ 0.50) | 0.19620 | 0.95890 | 76.27% |
| JSMA | 0.97520 | 0.00060 | -97.46% |
| SVM-attack | 0.20586 | 0.99795 | 79.209% |
| DT-attack | 0.33895 | 0.00730 | -33.165% |

**Table 5.**
BFF-IDS Augmentation using FGSM 0.36, SVM, JSMA, DT

| Manipul. | BFF-IDS Detection Rate | BFF-IDS (AUG) Detection Rate | Recovered rate by the BFF-IDS(AUG) |
|---|---|---|---|
| FGSM (∈ 0.29) | 0.17919 | 0.96443 | 78.524% |
| FGSM (∈ 0.36) | 0.17860 | 0.96515 | 78.655% |
| FGSM (∈ 0.50) | 0.19620 | 0.95958 | 76.338% |
| JSMA | 0.97520 | 0.00175 | -97.35% |
| SVM-attack | 0.20586 | 0.99935 | 79.349% |
| DT-attack | 0.33895 | 0.90070 | 56.175% |

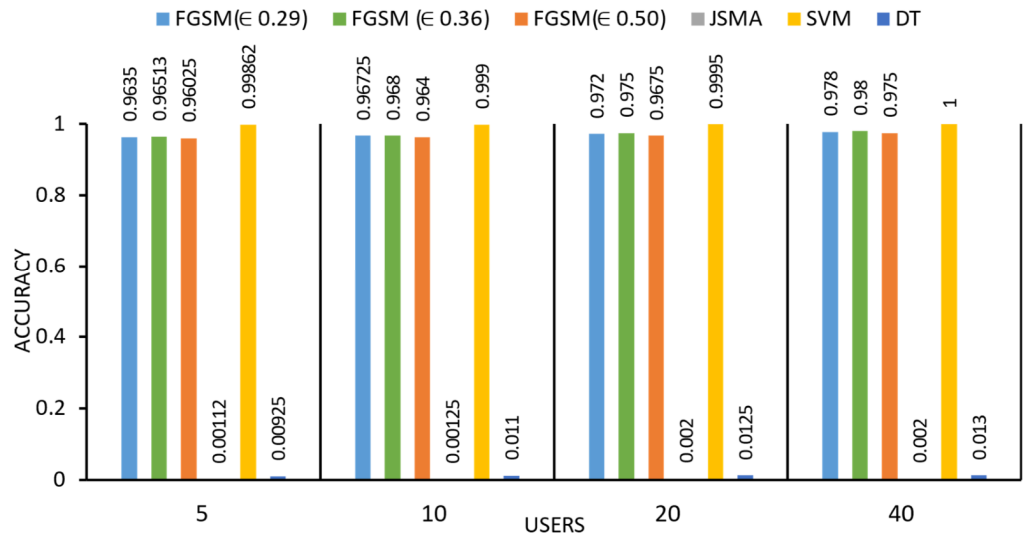

**Fig. 9.** BFF-IDS(AUG) Performance for Various Users using FGSM 0.36 and SVM for Augmentation

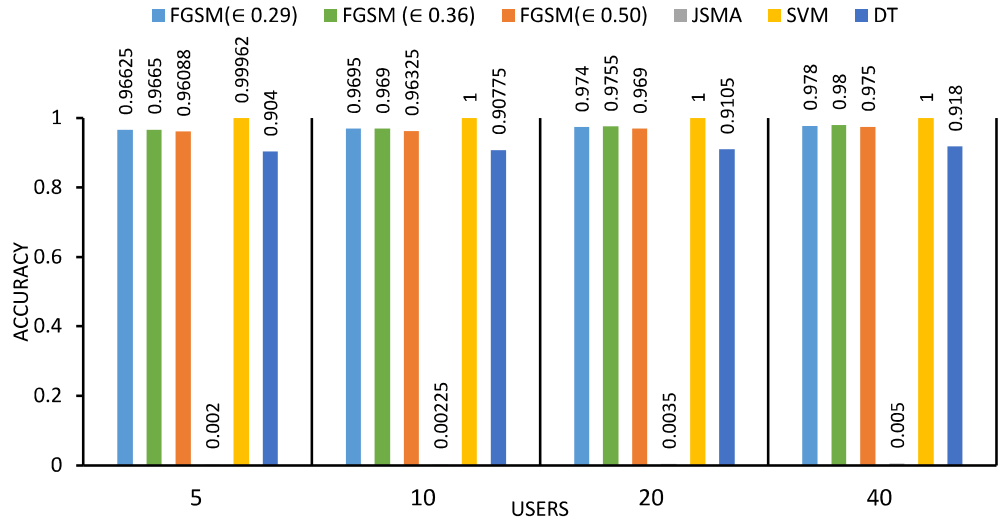

**Fig. 10.** BFF-IDS(AUG) Performance for Various Users using FGSM 0.36, SVM, JSMA and DT for Augmentation

*4.5 Robustness of the statistical test and BFF-IDS Augmentation*

*4.5.1 Robustness of the statistical test*

Section 4.3 assumes that the adversary can generate large adversarial samples, and the detector can conduct the statistical test on large sample sizes. This notion may not always be the case in practice, where the adversary is aware of the defense mechanism and decides to generate adversarial samples simultaneously using different methods to evade the BFF-IDS and the adversarial detector. As such, we investigate the confidence of our framework when presented with such a realistic situation.

We considered two scenarios to demonstrate the robustness of the statistical test as suggested by [16]. Firstly, we considered a situation where an adversary embeds adversarial samples generated from a single attack model in a benign sample. Each adversarial sample mixes with varying percentages of the benign sample (see Fig. 11). As shown in the results, the confidence of the statistical test decreases with an increase in the proportion of benign samples. The confidence in detecting FGSM (∈ 0.5) significantly degraded when the proportion of benign samples went to about 40%. For SVM, the confidence was affected at just beyond 10% of benign sample presents. Although FGSM (∈ 0.29) shows high $H_0$, the confidence equally degraded with an increase in the proportion of benign samples. JSMA and DT maintain high $H_0$ scores across all the mixtures. For this scenario, the $H_0$ acceptance amid benign samples is consistent with the findings in section 4.3 – the higher the MMD value of a technique the easier it is to be detected.

Secondly, we considered a scenario where the adversary simultaneously launches the attack using more than one adversarial crafting algorithm amid benign traffic. As indicated in Fig. 12, the confidence of the test equally degrades with an increase in the proportion of the benign sample. The mixture of

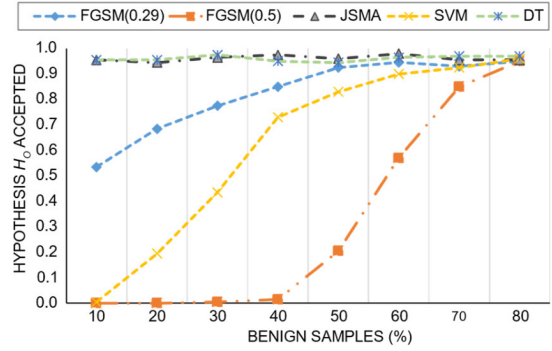

**Fig. 11.** Hypothesis *Ho* Test on Adversarial Examples and Benign Data.

FGSM (0.29)/SVM shows the highest acceptance score of 0.36 at 10% benign proportion, while the mixture of FGSM (0.5)/SVM maintains the rejection of the H0 from 10%-20% of the benign proportion. Consequently, the detection of adversarial examples mixed benign samples becomes difficult among a small set of inputs- the confidence of the statistical test degrades with a decrease in the proportion of adversarial samples present in the mixture of adversarial and benign samples.

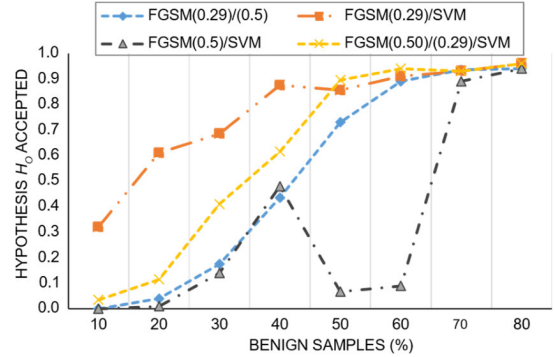

**Fig. 12.** Hypothesis *Ho* Test on a Different Mixture of Adversarial Examples and Benign Data.

*4.5.2 Robustness of BFF-IDS Augmentation*

Section 4.4 presented the BFF-IDS(AUG) performance in detecting the adversarial samples. In this section, we investigate the impact of the inclusion of adversarial samples class on the general performance of the model using the two training samples used in the augmentation as discussed in section 4.4.

Fig. 13 presents the impact of training the model with FGSM (0.36) and SVM as well as with FGSM (0.36), SVM, JSMA and DT as the augmented samples for the adversarial class. For the FGSM (0.36) and SVM, the general performance of the model was significantly high with 0.97667, 0.97607, 0.97619, and 0.97607 scored for precision, recall, f1-score and accuracy, respectively. The second sample with FGSM (0.36), SVM, JSMA and DT show a significant impact on the general performance of the model. The model performance dropped to about 0.72963, 0.63266, 0.63447 and 0.63266 for precision, recall, f1-score and accuracy, respectively. In all cases, the model's confidence improved compared to the model without augmentation, which recorded about 0.15927, 0.17860, 0.15779 and 0.17860 for precision, recall, f1-score and accuracy, respectively. Likewise, the confusion matrix shows how the adversarial example reduced the model's confidence by confusing the model to misclassify most of the samples as DOS attacks (see Fig. 14 (a)). Augmenting the model using all the samples significantly improved the performance. However, the model had difficulty classifying the adversarial samples correctly, attack-free from impersonation and impersonation from fuzzy attack (see Fig. 14 (b)). However, by removing those samples with lower MMD values, i.e., JSMA and DT, the model could generalize with more superior accuracy (see Fig. 14 (c)).

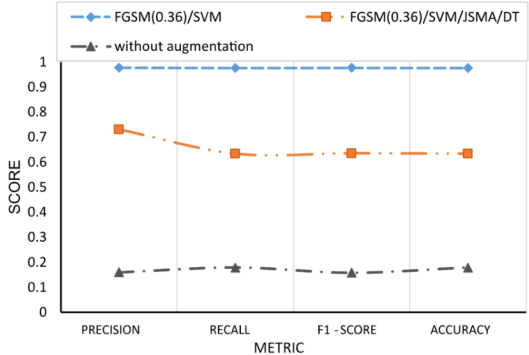

**Fig. 13.** Robustness of BFF-IDS (AUG) using Various Adversarial Examples.

## 5. Discussion

Adversarial examples significantly impact the confidence of BFF-IDS. In particular, FGSM with perturbation higher than 0.08, SVM attack and DT attack have the most impact on the model. These attacks degraded the model's accuracy from more than 0.975 to below 0.34. Among the adversarial examples investigated in this study, only JSMA non-targeted attack is found not to impact the model's confidence. This result is consistent with statistical divergence measures, MMD and ED, in section 4.1.

However, the question remains which metric between MMD and ED is a better indicator of how the BFF-IDS would be affected by an adversarial sample. Although the DT attack recorded a lower value for ED than benign data, it was still successful in degrading the performance of the BFF-IDS as its MMD value was higher than benign data. In the case of JSMA, the ED value was higher than benign data but still couldn't affect the model's performance as the MMD value was lower. Thus, the MMD is a better measure to detect adversarial examples/unknown attacks in BFF-IDS for CAN.

The results in section 4.3 show that the statistical hypothesis, $H_0$, effectively detects adversarial examples. Except for JSMA and DT attacks, a minimum of 50-100 samples are sufficient to detect most of the attacks in class-wise and mixed samples containing all classes of attack samples. However, it was discovered that mixed samples containing random classes of the attacks required more sample size than the class-wise for the statistical test to detect the adversarial samples. Consequently, there is a significant statistical distinction between benign and generated adversarial samples; thus, the statistical approach is sufficient to detect adversarial samples.

However, the statistical test cannot detect adversarial samples on a single-input basis and its confidence diminishes with a decrease in the number of samples in a batch. Thus, the augmentation of the BFF-IDS by retraining the model with detect samples of adversarial examples is effective in detecting adversarial examples per input. Except for JSMA and DT, the augmentation resulted in a recovery rate of more than 76.20% in both the augmentation scenario considered in section 4.4.

Furthermore, both the statistical hypothesis, $H_0$, and the BFF-IDS augmentation, BFF-IDS (AUG), are affected by the size and type of adversarial samples examined. As observed in section 4.5.1, the robustness of the hypothesis, $H_0$, is affected by the proportion of benign samples among adversarial samples- the confidence of the statistical test degrades with a decrease in the proportion of adversarial samples present in the mixture of adversarial samples and benign samples. On the other hand, the overall effectiveness of the BFF-IDS(AUG) is affected by the type of samples. Particularly, the addition of JSMA and DT-attack to the training samples significantly diminishes the model's overall performance (accuracy) from more than 0.97 to about 0.73. Consequently, the statistical test using MMD provides a good measure of which samples should be included for the augmentation. The higher the MMD value of the adversarial example from the benign sample, the better. In other words, the FGSM (<0.1), JSMA and DT-attack fail to achieve the goal of the adversary of reducing the confidence of the model and including them in training data significantly affects the accuracy of the BFF-IDS(AUG).

Although the statistical test and model augmentation approaches were motivated by [16], the main difference between our works is that we used the cross-technique transferability principle while they focused on inter-technique transferability. We investigated the impact of several adversarial examples crafting algorithms in which the adversary

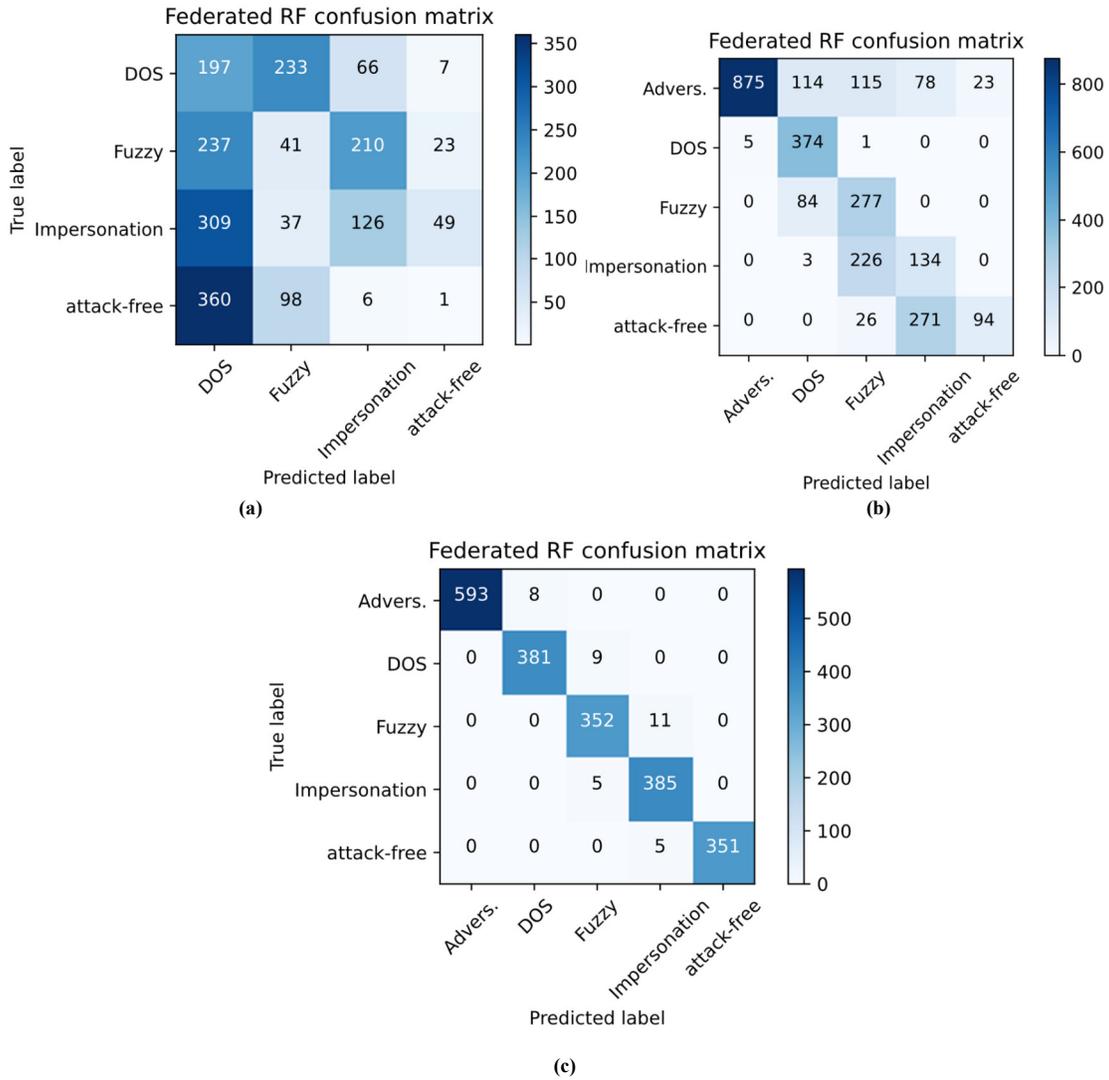

**Fig. 14.** Confusion matrix for BFF-IDS and BFF-IDS (AUG) (a) adversarial example (FGSM 0.39) attack on initial BFF-IDS (b) BFF-IDS (AUG) performance with FGSMA (0.39), SVM, JSMA and DT as augmentation data (c) BFF-IDS (AUG) performance with FGSMA (0.39) and SVM as augmentation data

has only knowledge of the features used to train the model. Also, their works focus on MNIST, DREBIN and MicroRNA datasets using the traditional learning approach, while we investigated on CAN dataset using BFF-IDS built by the FL concept. In addition, we proposed the continuous retraining of the model with the detected unknown samples in a sandbox environment to enable the detection of unforeseen adversarial examples.

Considering other related works, [23] demonstrated how ZOO and GAN attacks successfully degraded NIDS performance. However, there was no defense mechanism proposed against the attacks. Unlike our study, which offers mitigation measures, the proposed testbed in [27] facilitates the investigation of the impact of adversarial examples on IDS; the software offers no room for testing mitigation measures. Although the work of [28] is on CAV, the study focus only considered vehicular ad-hoc networks using synthetic datasets and binary classification problems.

On the other hand, we used real-world datasets while focusing on multi-class (five) classification problems, including the adversarial sample class.

To the best of our knowledge, the only available work that directly deals with adversarial examples on the CAN bus is [22]. The car hacking dataset (CARHD) used in that study was provided and probably collected under similar conditions by the same laboratory as ours, OTID. Also, the adversarial mitigation method, AADS, was similar to our approach. However, we utilized CAN ID cycles as features while they employed the raw data, which needed to be decoded from hexadecimal to decimal format, making their method more complex. In addition, their model was built using the traditional method, whose limitation was highlighted in the introduction section. Compared to the maximum perturbation $\epsilon$, 0.5 we considered in our experiments, the authors considered a very high perturbation $\epsilon$ of 5, which may be forbidden in the CAN bus specification-High perturbations are mostly acceptable in image tasks. Finally, their proposed system has no mechanism to detect unknown adversarial examples and, therefore, is unsustainable in real-life deployment. As presented in Fig. 15, our proposed model relatively shows competitive results despite using the difficult features, transferability principle, and more attack classes.

## 6. Conclusion

In this work, we set out to establish the vulnerability of BFF-IDS to adversarial examples, detect the adversarial examples and augment the BFF-IDS to detect such examples and make it more resilient. We investigated the vulnerability of BFF-IDS by relying on the adversarial sample transferability to investigate the impact of several adversarial sample algorithms, including FGSM, JSMA, SVM-attack, and DT-attack. We relied on a threat model to determine whether an adversary knows the features needed to significantly diminish the confidence of the BFF-IDS using an adversarial system. To augment BFF-IDS to detect adversarial examples, we relied on the notion that generated samples may have a different statistical distribution from benign samples. Resilience is protected by augmenting BFF-IDS with an additional class for the detected samples, which allows for detecting single adversarial examples. We demonstrate the robustness of the statistical test by considering the mixture of adversarial samples from various algorithms and benign samples. Also, we investigated the robustness of the BFF-IDS augmentation by investigating how the mixture of several combinations of adversarial examples from the different algorithms as training samples can affect the detection of the adversarial samples and the general performance of the model.

The most important results of this research are the following:

- BFF-IDS is very vulnerable to adversarial examples attacks as it succeeded in significantly reducing the confidence (accuracy) of our model from more than 97% to as low as 20% in some instances.
- MMD is an effective statistical measure for detecting adversarial examples and unknown samples. The statistical detector integrated with BFF-IDS effectively detects adversarial samples from benign sample distributions in large batch samples. However, the statistical test cannot identify which samples are adversarial.
- We found that BFF-IDS, when augmented with the class of detected examples, significantly improves the confidence and thus resilience of the model, as it allows for detecting attacks on a single input basis.

The BFF-IDS(AUG) performance significantly depends on the combination of adversarial samples included for the argumentation. In particular, adversarial samples with higher values of MMD when compared to benign data are more suitable as they do not diminish the overall accuracy of the model.

Therefore, we conclude that the proposed integration of the statistical test as an adversarial detector and the subsequent augmentation of the BFF-IDS with detected adversarial samples provides a sustainable security framework against adversarial examples and other unknown attacks. In the bigger picture, this method helps deriving benefits from the huge (big) data generated in the smart city by detecting unknown samples and using them to augment security models and provide other room to analyze unforeseen novel attacks.

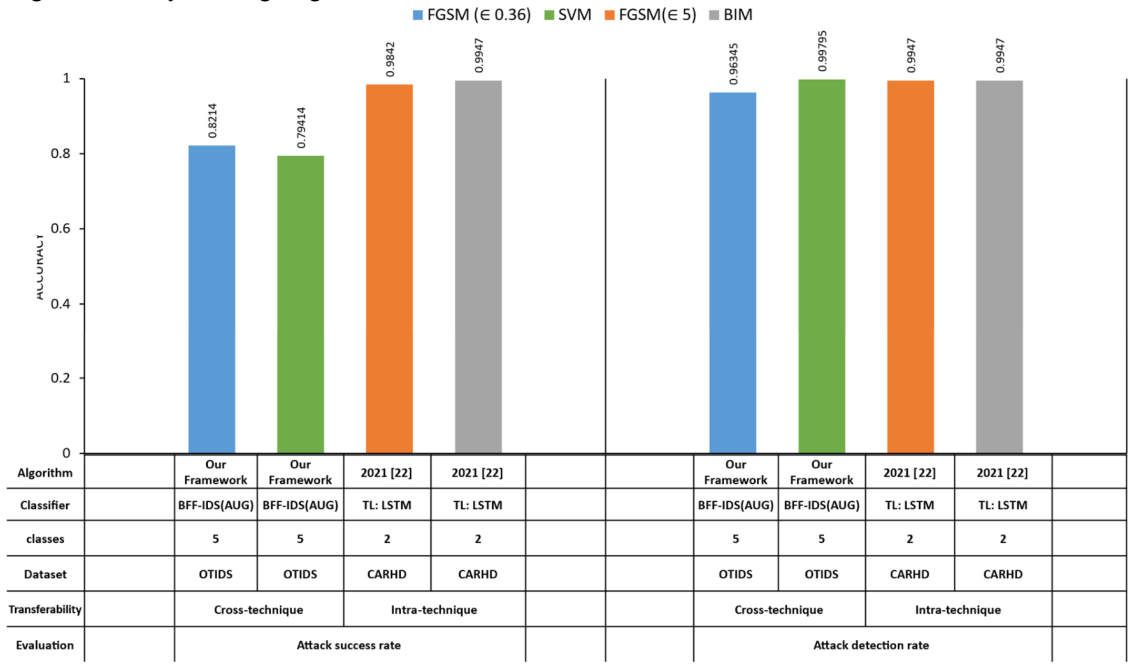

**Fig. 15.** Performance evaluation of BFF-IDS(AUG) against other works

Further studies are needed to establish the acceptable perturbation based on the CAN protocol. Also, a black-box scenario where the adversary has no knowledge of the feature but can generate CAN traffic should be investigated.

**Declaration of Competing Interest**

The authors declare that they have no known competing financial interests or personal relationships that could have appeared to influence the work reported in this paper.

**Acknowledgements**

This research is supported by Jeollannam-do (2021 R&D supporting program operated by Jeonnam Technopark) and financially supported by the Ministry of Trade, Industry and Energy (MOTIE) and Korea Institute for Advancement of Technology (KIAT) under the research project: "National Innovation Cluster R&D program" (Grant number: 1415175592 & P0016223). It is also supported by Institute of Information & communications Technology Planning & Evaluation (IITP) grant funded by the Korean government (MSIT) (No. 2021-0-02068, Artificial Intelligence Innovation Hub). We acknowledge the authors, K. Grosse et al., for granting us access to their source code to conduct our investigations.